\RequirePackage{fix-cm}
\documentclass[twocolumn]{svjour3}          
\smartqed  
\usepackage{graphicx}
\usepackage{amssymb}
\usepackage{color}
\usepackage{amsmath} 

\begin{document}

\title{Partially unstable attractors in networks of forced
integrate-and-fire oscillators
}



\author{Hai-Lin Zou         \and
        Zi-Chen Deng \and  
        Wei-Peng Hu \and
        Kazuyuki Aihara  \and
        Ying-Cheng Lai
}


\institute{Hai-Lin Zou \at
              School of Mechanics, Civil Engineering and Architecture, Northwestern Polytechnical University, Xi’an 710072, China\\
              \email{zouhailin@nwpu.edu.cn}           
           \and
           Zi-Chen Deng \at
              School of Mechanics, Civil Engineering and Architecture, Northwestern Polytechnical University, Xi’an 710072, China
           \and
           Wei-Peng Hu \at
              School of Mechanics, Civil Engineering and Architecture, Northwestern Polytechnical University, Xi’an 710072, China
           \and
           Kazuyuki Aihara \at
              Institute of Industrial Science, University of Tokyo, 4-6-1 Komaba, Meguro-ku, Tokyo 153-8505, Japan
           \and
           Ying-Cheng Lai \at
              School of Electrical, Computer and Energy Engineering,
              Arizona State University, Tempe, AZ 85287, USA
}

\date{Received: date / Accepted: date}

\maketitle

\begin{abstract}
The asymptotic attractors of a nonlinear dynamical system play a key role
in the long-term physically observable behaviors of the system. The study
of attractors and the search for distinct types of attractor have been a
central task in nonlinear dynamics. In smooth dynamical systems, an
attractor  is often enclosed completely in its basin of attraction with
a finite distance from the basin boundary.  Recent works have uncovered that,
in neuronal networks, unstable attractors with a remote basin can arise,
where almost every point on the attractor is locally transversely repelling.
Herewith we report our discovery of a class of attractors: partially unstable
attractors, in pulse-coupled integrate-and-fire networks subject to a
periodic forcing. The defining feature of such an attractor is that it can
simultaneously possess locally stable and unstable sets, both of positive
measure. Exploiting the structure of the key dynamical events in the
network, we develop a symbolic analysis that can fully explain the emergence
of the partially unstable attractors. To our knowledge, such
exotic attractors have not been reported previously, and we expect them
to arise commonly in biological networks whose dynamics are governed by
pulse (or spike) generation.
\end{abstract}

\section{Introduction}

A variety of physical, biological, and chemical processes can be described
by dissipative dynamical systems, for which the asymptotic behaviors are
determined by attractors - a fundamental class of dynamical invariant
sets. The concept of attractor plays a pivotal role in
the development of nonlinear dynamics and chaos theory~\cite{Ott:book}, and
it is also important for understanding many fundamental phenomena in nature.
For example, the computational capability for  neural networks is determined
essentially by the attractors~\cite{HOPFIELD:1982}. Multiple attractors may coexist, where an essential goal is to analyze the related global dynamics~\cite{ZHANG:2015,liu2016global}.

Attractors in nonlinear dynamical systems are often asymptotically stable
in the sense that they ``attract'' nearby initial conditions.   Associated with an attractor is its basin of attraction, where an initial condition in this region leads to a trajectory that approaches the
attractor asymptotically.   There can be another type of attractors for which does not require the local stability. Mathematically, such
an attractor can be defined in terms of its basin measure that must still
be positive in order for it to attract initial conditions. These are the
Milnor attractors~\cite{M:1985}, which are typically chaotic, possess locally
unstable directions, and have strength zero~\cite{K:1997}.
One class of Milnor
attractors are those with a riddled basin~\cite{AYYK:1992,OSAKY:1993,ABS:1994,HCP:1994,LGYV:1996,LG:1996,NU:1996,ABS:1996,Lai:1997,BCP:1997,MKS:1997,MMPM:1998,KMSB:1998,LG:1999b,WM:2000,LT:book,munteanu2014chaos}.
Specifically, for such an attractor, there exists a set of measure-zero
points on it with transversely unstable dynamics. Because of the ``repulsion''
in the transverse direction, infinitesimally away from the attractor there
is a set of positive measure, initial conditions
from which approach asymptotically some different, coexisting attractor in
the phase space. As a result, for any initial condition attracted to the
Milnor attractor, there are initial conditions arbitrarily nearby that
generate trajectories towards another attractor. The basin of the Milnor
attractor is thus riddled with ``holes'' that belong to the basin of the
other attractor, hence the term riddled basins~\cite{AYYK:1992,OSAKY:1993,ABS:1994,HCP:1994,LGYV:1996,LG:1996,NU:1996,ABS:1996,Lai:1997,BCP:1997,MKS:1997,MMPM:1998,KMSB:1998,LG:1999b,WM:2000,LT:book,munteanu2014chaos}.

A different type of Minor attractors is {\em unstable attractors}, which
constitute locally unstable saddles but with a ``remote'' basin of positive
measure~\cite{TWG:2002a,TWG:2003a}. Such attractors are ubiquitous in
a generic class of biological systems: networks of pulse-coupled
integrate-and-fire oscillators, which have been studied extensively
for the collective
dynamics of biologically realistic networks~\cite{MS:1990,V:1996,GV:2007,KA:2010,LP:2010,LOPT:2012}. For example, such model has been widely used to study the emergence of irregular
states \cite{ZTGW:2004,ZLPT:2006a,JMT:2008,KT:2009a,ZBH:2009,ZGC:2009,RL:2014}, and the stabilities of
various dynamical states~\cite{TWG:2002,AS:2010,OPT:2014,AM:2015}.
For the  unstable attractors, there are generic dynamical events
that can make the phase differences among the oscillators grow, effectively
driving the system away from the unstable attractor~\cite{zou:2010}.
The existence of unstable attractors in pulse-coupled integrate-and-fire
oscillator networks has been established for four~\cite{AT:2005} and an
arbitrary number of globally coupled oscillators~\cite{BES:2008size}.

An advantage of unstable attractors, due to their
unstable local dynamics, is that they can be exploited for control and
information processing. In particular, points arbitrarily close to an
unstable attractor can approach another unstable attractor, forming
heteroclinic connections~\cite{BES:2008,KM:2008,KGM:2009}. As a result,
switching among the attractors can occur following the natural dynamical
evolution, into which information reflecting the input signal can be
encoded~\cite{NT:2012}. If the system possesses a large number of unstable
attractors, they can form a complex network through heteroclinic connections
in the phase space, which can be exploited for complex logical
computation~\cite{NT:2009,NT:2012}. In addition, the unstable attractors
can display certain metastable phenomena and be used to identify the
input driven dynamics~\cite{MPJ:2012}.

The unstable attractors reported in previous works all have a common
feature: their local dynamics are purely unstable. In this paper, we
report our finding of a novel class of  attractors: they possess
both unstable and stable local dynamics. In particular,
we investigate systems of pulse-coupled integrate-and-fire oscillators
subject  to a periodic driving force~\cite{KHR:1981} and uncover
attractors that exhibit two distinct types of response to perturbation:
stable and unstable. We name the attractors {\em partially unstable
attractors}. There is a key difference between the partially unstable attractors
reported in this paper and the attractors with a riddled basin: for the
former the set of unstable local points on the attractors
has a finite measure while for the latter, the set has measure
zero. We shall demonstrate that, the dynamical origin of the partially unstable
attractors can be fully understood through analyzing the dynamical
events. In addition to being fundamental to the dynamics of
integrate-and-fire networks, partially unstable attractors can be
advantageous from the standpoint of control, as the existence of both
locally unstable and stable dynamics offers richer possibilities for
control.

This paper is organized as follows: In the second section, we describe the model, and the simulation method for the system; in the third section, we show  examples of partially unstable attractors; in the fourth section, we investigate the effect of parameters on the emergence of partially unstable attractors and show that  these attractors exist in  systems   with various sizes or different coupling topologies; in the fifth section, the symbolic  events are used to understanding these attractors; finally, the conclusion and discussion are drawn.

\section{Networks of pulse-coupled integrate-and-fire oscillators subject  to periodic force} \label{sec:results}

\subsection{Description of the model}
Networks of pulse-coupled integrate-and-fire oscillators arise commonly
in neuronal systems where, for example, each unit can be a particular type
of neuron~\cite{TWG:2003a}. The dynamics of an individual integrate-and-fire
oscillator is equivalent to that of a phase oscillator, and the phase
variable can be obtained through a nonlinear transformation~\cite{TWG:2003a,MS:1990}. Previous studies focused on the case where the driving or
stimulation to each oscillator is constant so that the resulting attractors
are usually of  period one (with respect to an arbitrarily chosen
reference oscillator)~\cite{TWG:2002a,zou:2010,AT:2005,BES:2008size}.
In realistic situations more complicated driving can be expected.
For example, in biological systems periodic forcing is common.
We thus set out to investigate the dynamics of networks of integrate-and-fire
oscillators subject to a periodic forcing~\cite{KHR:1981}. In this
case, periodic attractors with various periods can be generated.
Mathematically, such a system of $N$ oscillators can be described by
\begin{equation} \label{EqSys}
\frac{d{V}_{i}}{d{t}}= - \gamma V_{i}+I+B\cos{(\omega t)}+\sum_{j=1}^{N}
\sum_{k\in \mathbb{Z}+} \varepsilon_{i,j}\delta(t-t^{s}_{j,k}-\tau),
\end{equation}
where $V_i$ denotes the state of oscillator $i$ and
 $\gamma$ accounts for the dissipation or leaky effect~\cite{KHR:1981}, which
is fixed to be unity in our study.
The parameter $I$ is the constant
bias of the applied current, $B\cos{(\omega t)}$ is the external
periodic driving current of frequency $\omega$ and amplitude $B$, and
the function $\delta(\cdot)$ is the Dirac delta function. When the state
of the $i$th oscillator reaches a threshold (conveniently set to unity),
its state is reset to zero and a pulse is generated. The pulse will be
received, after a time delay $\tau$, by other oscillators that have
incoming links from oscillator $i$. The summation term on the right-hand
side of Eq. \ref{EqSys} accounts for the action of pulses from other
oscillators on oscillator $i$, where $\varepsilon_{i,j}$ is the normalized
coupling strength from oscillator $j$ to oscillator $i$:
$\varepsilon_{i,j}=\varepsilon/k_i$ with $k_i$ being the number of
incoming links of oscillator $i$. The normalization is essential to
the study of network collective dynamics such as
synchronization~\cite{TWG:2002}. For a globally coupled network, any
two oscillators are coupled but self-links are excluded. Thus we have
$\varepsilon_{i,j}=\varepsilon/(N-1)$ for $i\neq j$. To be concrete,
in this paper we consider excitatory coupling, i.e., $\varepsilon>0$.

\subsection{Dynamics of free evolution.}
For each oscillator, there are two distinct dynamical events: it can
generate pulses when its state reaches the threshold or receive pulses
from other oscillators. Between two successive events, the state $V_i$
of oscillator $i$ evolves freely according to
\begin{equation} \label{EqSingle}
\frac{d{V_i}}{d{t}}=I+B\cos{(\omega t)} - V_i,
\end{equation}
the solution of which is
\begin{equation} \label{Sol_c}
V_i(t)=C e^{-t}+\frac{B[ \omega \sin{(\omega t)} +
\cos{(\omega t)}]}{1+\omega^2 }+I.
\end{equation}
Here $C$ is a parameter determined by a specific  initial condition. Suppose that
the state of oscillator $i$ at time $t_a$ is $V_i^a<1$. With this initial
condition, we can get an explicit solution of Eq.~\ref{Sol_c}. For
convenience, we define a function $F(t,t_a,V_i^a)$:
\begin{eqnarray}
F(t,t_a,V_i^a) = & \frac{B \omega \sin{(\omega t)} + B\cos{(\omega t)}}{\omega^2+1} \nonumber\\
& + I - e^{t_a-t} \{ -V_i^a  \nonumber\\
&+ \frac{B \omega \sin{(\omega t_a)} + B \cos{(\omega t_a)}}{\omega^2+1}+I,  \label{eq7} \nonumber
\end{eqnarray}
so that the solution under the initial condition $(t_a,V_i^a)$ can be
expressed as
\begin{equation} \label{eq_solSingle}
V_i(t)=F(t,t_a,V_i^a).
\end{equation}
The solution will be used in simulating the system dynamics.


\subsection{Firing time and simulation of the system}

The dynamics of the pulse-coupled integrate-and-fire system is composed
of free evolution, which entails integrating Eq.~(\ref{EqSingle}), and
frequent disturbances from dynamical events such as firing and arrival
of pulses. It is essential to determine the firing time $t_f$ due to the
free evolution through Eq.~\ref{eq_solSingle}. To do so one first
identifies a time interval with two end points, at which the states
values are smaller and larger than the threshold, respectively. The
underlying oscillator, say $i$, can fire during this interval due to
the free evolution. One then applies a bisection technique to
systematically reduce the length of the time interval to locate the
firing time accurately.

A simple bisection algorithm is as follows. For oscillator $i$ with
state $V_i^a<1$ at $t_a$, one first advances the state according to
Eq.~\ref{eq_solSingle} with a small initial time step $h_0=0.01$
until the state value exceeds unity, say at time $t_R$. Let $t_L=t_R-h_0$
denote the time when the corresponding state is below unity.
The true firing time is within the interval $[t_L,t_R]$. One then applies
bisection to gradually reduce the time interval $[t_L,t_R]$. Specifically,
one obtains the value $V_i(t_M)=F(t_M,t_a,V_i^a)$ for the middle time
$t_M=(t_L+t_R)/2$. If the state is below unity, one updates $t_L=t_M$.
Otherwise, one has $t_R=t_M$. The length of the interval becomes
$h_1=h_0/2$. This bisection process can be applied repetitively until
the length of the interval $t_R-t_L$ is smaller than a small precision
value, e.g., $10^{-14}$. The final time $t_R$ is recorded as the next
firing time $t_f$. For a network of oscillators, the oscillator with
the largest state value can be chosen to determine $t_f$.

Simulation of the system can be described as follows. The firing and
arrival of pulses are the two types of events that interrupt the free
evolution. It is convenient to use a vector $V(t)$ to represent the
states of all oscillators at time $t$.

\begin{description}
\item[\emph{Step 1}] Determine the next firing time $t_f$ for the
system due to free evolution. The oscillator with the largest
state will reach the threshold first due to the free evolution. Then
choose the oscillator with the largest state at time $t$ and determine
the next firing time $t_f$ using the bisection technique.

\item[\emph{Step 2} ] Compare $t_f$ with the next time $t_r$ of pulse
receipt. If $t_f<t_r$, go to Step 3, otherwise go to Step 4. In the
case where no pulse is waiting to be processed, which can occur when all
pulses have been received or no pulse has been generated yet, go to Step 3.

\item[\emph{Step 3}] The next event is that one or more oscillators fire
at time $t_f$, given that the state of oscillator $i$ is $V_i(t)$ at
time $t$, for $i=1,2,\ldots,N$. Update the state at $t_f$ using
$F(t_f,t,V_i(t))$ for each oscillator $i$, record the pulses when
the state of an oscillator reaches the threshold, and reset the state
of the corresponding oscillator to zero. Finally the time $t$ is
updated to $t_f$.

 \item[\emph{Step 4}] The next event is the arrival of pulses at time
$t_r$, given that the state of oscillator $i$ is $V_i(t)$ at time $t$,
for $i=1,2,\ldots,N$. First update the state at time $t_r$ using
$F(t_r,t,V_i(t))$ for each oscillator $i$. Calculate the increment $E_i$
for the state of each oscillator $i$. Here
$E_i=\sum_{k\in A} \varepsilon_{ik}$, where $A$ denotes the set of
oscillators whose pulses are received at the current time $t_r$. Then
update the states of oscillators with the increments, such as $V_i=V_i+E_i$
for  oscillator $i$. When some oscillators reach the threshold, record
the pulses and reset the states of these oscillators to zero. Update
the time to $t=t_r$.
\end{description}

\section{Partially unstable attractors}

\subsection{Attractors under the return map}
Any periodic attractor of system (\ref{EqSys}) lives in a high dimensional
phase space with an uncountably infinite number of points (versus a steady
state that contains a single point). To analyze a large number of periodic
attractors directly is challenging, but the method of Poincar\'{e} surface
of section provides an effective way of
probing into the dynamics~\cite{Ott:book}. Specifically, we monitor the state
of the whole coupled system when a reference oscillator, e.g., oscillator 1,
resets itself, effectively generating a Poincar\'{e} map or, equivalently, a
{\em return map}. The map evolves the state of the system right after
the reset of the reference oscillator to that of the system immediately
after the next reset.
Under the return  map, a periodic attractor is
composed of a finite number of points, where each point corresponds to
the states of oscillators of the form $(0,V_2,\ldots,V_i,\ldots,V_N)$,
with $V_i$ being the state of oscillator $i$ ($V_1 = 0$ because oscillator
1 is the reference oscillator). The return map approach has been widely
used to analyze the dynamics of general pulse-coupled
systems~\cite{TWG:2002a,TWG:2003a,AT:2005,BES:2008,zou:2010,KM:2008}.

Under the return map, the attractors of the system can be assessed through
the local dynamics of the points that constitute each attractor.
For each point, we examine the trajectories starting from a small
neighborhood about the point to determine the Lyapunov stability. For
example, a period-one attractor corresponds to a single point on the
Poincar\'{e} surface of section. If there exists a neighborhood from
which almost all the trajectories starting diverge, this point is
\emph{unstable}. An unstable fixed point with a positive measure of the
basin is a period-one unstable attractor~\cite{TWG:2002a}. On the contrary,
if there exists a neighborhood such that all the trajectories originated
from it stays inside asymptotically, this point is \emph{stable}. Note that,
a periodic attractor of period-$M$ corresponds to a sequence of $M$ points
under the return map, repeating themselves after every $M$ resets of the
reference oscillator. It is equivalent to a fixed point of the $M$th
iterated map.

\subsection{Locating Partially unstable attractors}
We describe how to locate partially unstable attractors numerically following the  Lyapunov criterion.
One needs to monitor
the dynamical evolution of random instantaneous perturbation to a point within an attractor.  The perturbation can be generated in two steps.
First, the perturbed value $\tilde{\delta}_i$ for oscillator $i$ is
randomly chosen from the interval $\tilde{\delta}_i \in [-0.5,0.5]$.
Second, we rescale the perturbation as
\begin{equation} \label{eq_rescale}
\delta_i= D \tilde{\delta}_i /\sum_{i=1}^N |\tilde{\delta}_i|,
\end{equation}
where $D$ is the strength of the perturbation.  Generally, $D$ should be
larger than the uncertainty in determining the firing time, which is on
the order of $10^{-14}$. We  then vary the value
of $D$ from, say, $10^{-13}$ to $10^{-10}$.
 After applying the instantaneous perturbation, the initial
distance between the perturbed and the original trajectories is
$d_0=\sum_i^N|\delta_i|=D$. To determine the stability, it is necessary
to monitor the dynamical evolution of the distance.

Specifically, for a point $P$ within an attractor of
period $M$, one requires computation
of the trajectories starting from the neighborhood of $P$. A neighboring point
$X$ can be chosen as $X_i=P_i+\delta_i$ for $i=1, 2, \ldots, N$. Starting
from $X$, one calculates a trajectory of length $L=500$, where each point
on the trajectory is the state of the system at every $M$ resets of the
reference oscillator. The initial distance is $d_0=\sum_i^N|\delta_i|$,
where the distance for any two points $Y$ and $Z$ is defined as
$d=\sum_i^N|Y_i-Z_i|$ and $d_0=D$ according to the rescaling process
[see Eq.(\ref{eq_rescale})]. One then measures the distance between each
point on the trajectory and $P$. This allows one to investigate whether
the distance is always smaller than $d_0$ or at some time exceeds $d_0$.

To determine the local stability of point $P$, one randomly
chooses a number (e.g., $n_L=30$) of neighboring points, and record the
states after every $M$ resets of the reference oscillator 1. This
generates $n_L$ trajectories. If all the tested trajectories stay within
$P$'s neighborhood, it is \emph{stable}. However, if all the trajectories
leaves this neighborhood eventually, the point is is \emph{unstable}.
Repeating this procedure, one can locate all the locally stable and
unstable points for the attractor. A simple criterion to locate the
partially unstable attractors is according to its definition: a
partially unstable attractor has at least one unstable point, while
other points are stable.

\subsection{Emergence of partially unstable attractors}
The emergence of partially unstable attractors is counterintuitive. In
particular, for a smooth dynamical system, the points belonging to a
periodic attractor of period-$M$ possess the same stability because
they all correspond to exactly the same fixed point of the $M$th
iterated map. What is then the difference between an unstable attractor
and a partially unstable attractor? By definition, almost all trajectories
from a neighborhood of an unstable attractor diverge from it, excluding
the possibility of existence of any stable point.

\begin{figure}
\centering
\includegraphics[width=\linewidth]{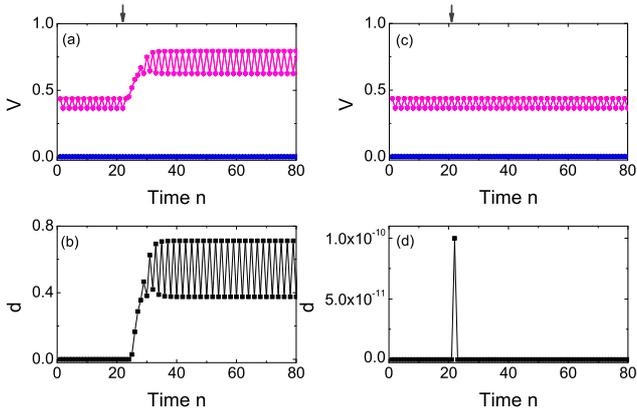}
\caption{
{\bf Responses of the instantaneous perturbation applied separately to
an unstable and a stable point of a period-2 partially unstable attractor.}
Perturbation of strength $D=10^{-10}$ is applied at the position indicated
by the arrow. In panels (a) and (c), the states $V$ of all oscillators
versus the discrete time $n$ (corresponding to the $n$th reset of the
reference oscillator 1) are shown. The state evolution of
  oscillators 1 and 3 is represented by the lower curve near zero, while
the upper curve is for the evolution of oscillators 2 and 4.
(a) Instantaneous perturbation on one point of the attractor can make
the system approach a new attractor (see text for a detailed explanation).
(b) The sequence of distance $d$ for the trajectory in (a) to the partially
unstable attractor. (c) The stable response of the instantaneous
perturbation on another point of the attractor. (d) The corresponding
sequence of distance for the trajectory in (c) to the partially unstable
attractor. The parameters of the system are $N=4$, $\tau=0.14$,
$\varepsilon=0.3$, $B=1.6$, $I=3$, and $\omega=10$.}
\label{fig_pun}
\end{figure}

\paragraph*{A simple example.} The simplest partially unstable attractor has period two, where one point
is unstable and another is stable. Figure~\ref{fig_pun} shows an
example of such an attractor in a system of $N=4$ globally coupled
integrate-and-fire oscillators. After the system settles into this attractor,
we let the system evolve without perturbation for $L$, say 10,
periods (corresponding to $2L$ resets of the reference oscillator). We then introduce instantaneous perturbation separately to the states of oscillators just after $2L+1$ and those just after $2L+2$.
Perturbation to the states of oscillators just after the 22nd (i.e., $L=10$) reset of reference oscillator 1 makes the
system approach a new attractor, as shown in Fig.~\ref{fig_pun}(a). This means that the corresponding point of the attractor is an unstable point.
 In order to show such process intuitively, we also
measure the distance between the point on the perturbed trajectory
to the original attractor. For example, the $i$th point on the
perturbed trajectory is denoted by $P_i$. An attractor with
period $m$ is represented by $m$ points, where each point is denoted
as $Q_i$. The distance between point $P_i$ and the attractor is $
d=\min_j \sum_i^N|P_i(k)-Q_j(k)|$.
Here $P_i(k)$ and $Q_j(k)$ denote the state of oscillator $k$ for
the $i$th point on the perturbed trajectory and the $j$th point
on the attractor, respectively.
The distance between the corresponding point on the perturbed
trajectory and the original attractor,  is shown in
Fig.~\ref{fig_pun}(b), where the distance finally becomes large,
signifying the unstable nature of this point. The attractor, however, is
stable with respect to perturbation on the other periodic point, as
shown in Fig.~\ref{fig_pun}(c), with the corresponding distance sequence
displayed in Fig.~\ref{fig_pun}(d). The system deviation from the periodic
point due to the perturbation at the position of the arrow becomes zero
after one reset of the reference oscillator, indicating the stable nature
of the point. Physically, this is due to the passive firing on which the
perturbation has little effect.

\begin{figure}
\centering
\includegraphics[width=\linewidth]{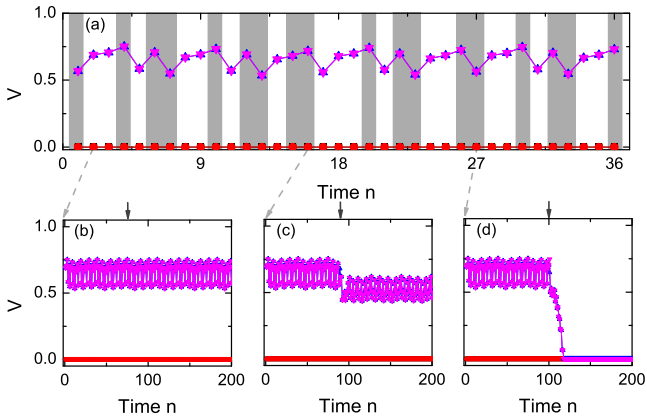}
\caption{ {\bf A partially unstable attractor of period 36}.
(a) Temporal evolution of all oscillators associated with a partially
unstable attractor of period 36, where the unstable points are highlighted
in gray, and the remaining points are stable. (b) The stable response of
the system to perturbation applied at the position of the arrow.
(c-d) Perturbation to the unstable point can make the system approach
different attractors. Here time is represented by the $n$th reset of the
reference oscillator 1, and the strength of the perturbation $D=10^{-12}$.
System parameters are $N=4$, $\tau=0.2$, $\varepsilon=0.3$, $B=1.48$, $I=3$,
and $\omega=10$.}
\label{fig_long_pua}
\end{figure}

\paragraph*{A complex example.} For a partially unstable attractor with a period larger than two, the
numbers of stable and unstable points can vary, and these two types of
points can be mixed in a complicated way. Figure~\ref{fig_long_pua} shows
a partially unstable attractor of period 36 in a system of $N=4$ globally
coupled oscillators, where the unstable points are highlighted in gray (the
remaining points are stable). The system state is robust against perturbation
to the stable points, as shown in Fig.~\ref{fig_long_pua}(b). However,
perturbation to the unstable points can make the system transition to
new attractors, as shown in Figs.~\ref{fig_long_pua}(c,d).

To locate the partially unstable attractors with long periods can be
extremely computationally demanding, as it is necessary to examine each
point's local dynamics. It has been known that periodic orbits of long
periods are typical in the forced integrate-and-fire
oscillators~\cite{KHR:1981}, and even quasi-periodic attractors can
emerge.  Thus we focus on partially unstable attractors
of period $2$ to address the issues of their existence and dynamical origin.

\begin{figure}
\centering
\includegraphics[width=\linewidth]{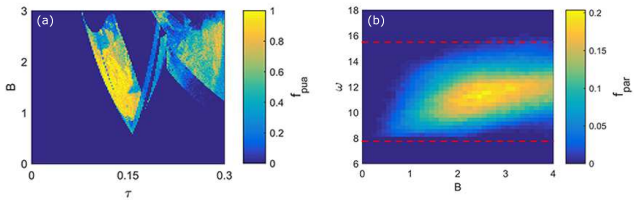}
\caption{ {\bf Parameter dependence of partially unstable attractors}.
For a system of $N=4$ oscillators, the dependence of the probability
of partially unstable attractors of period two on parameters $\tau$ and
$B$. (a) The fraction of initial conditions, $f_{pua}$, that lead to these
attractors. The number of random initial conditions for each pair of
parameters, $(\tau, B)$, is 200. (b) Dependence of $f_{par}$ on parameters
$B$ and $\omega$, where $f_{par}$ is the relative size of the parameter
region in $(\tau,\varepsilon)$ with period-2 partially unstable attractors.
The two dashed lines indicate the estimated region for $\omega$. Other
parameters are $\varepsilon=0.3$, $I=3$, and $\omega=10$. The strength $D$
of the perturbation is randomly chosen from the range
[$10^{-13}$, $10^{-10}$].}
\label{fig_para}
\end{figure}

\section{Typicality and robustness of partially unstable attractors.}
\subsection{Partially unstable attractors under various parameters}
We first identify the regions in the parameter plane ($\tau,B$) in which
period-2 partially unstable attractors arise, for fixed $\omega=10$, $I=3$,
$\varepsilon=0.3$, by calculating the fraction $f_{pua}$ of initial
conditions that lead to these attractors.
The $f_{pua}$ is obtained by using
200 random initial conditions.
The results are shown in Fig.~\ref{fig_para}(a).
We see that, in the parameter plane, the probability for generating period-2
partially unstable attractors is appreciable.

We next turn to the parameter $\omega$, the frequency of the
applied current that defines an external time scale. An individual oscillator
has its own time scale associated with its local dynamics, i.e., the time
of free evolution from the reset to the next firing. It is useful to determine
the relation between the two time scales, which can be done by analyzing
the period of the  free dynamics. Suppose that oscillator $i$ resets
at $t_a$ with state $V_i^a=0$. The dynamics of free evolution can be obtained
from Eq.~\ref{eq_solSingle} as
\begin{equation}\label{eq_vit}
\begin{split}
  V_i(t) =  & \frac{B \omega \sin{(\omega t)} + B\cos{(\omega t)}}{\omega^2+1}+ \\
    & I-e^{t_a-t}\{ \frac{B \omega \sin{(\omega t_a)} + B \cos{(\omega t_a)}}{\omega^2+1}+I\}.
\end{split}
\end{equation}
The time duration $T_d$ before the threshold is reached again is
$T_d=t-t_a$ for $V_i(t)=1$. The first term on the right-hand side of
Eq.~\ref{eq_vit} can be written as
\begin{equation}
\frac{B \omega \sin{(\omega t)} + B\cos{(\omega t)}}{\omega^2+1}
= \frac{B}{\sqrt{\omega^2+1}} \sin{(\omega t+ \theta)} ,
\end{equation}
where $\theta=\arctan{(1/\omega)}$. The maximum value of this term is about
$B/\omega$. The effect of this term on $T_d$ can be neglected, if $B/\omega$
is much smaller than $I$,  leading to $V_i(t) \approx I-e^{t_a-t}I$.
We thus have
\begin{equation}
T_d \approx \log{(\frac{I}{I-1})}.
\end{equation}
For the applied current with the frequency $\omega$, the time scale is
$T_\omega=2\pi/\omega$. We can then investigate the interplay between $T_d$
and $T_\omega$ for period-2 partially unstable attractors, where each
oscillator resets itself two times during $T_\omega$ time duration. Since $T_d$
only takes into account the free evolution and the arrival of pulses can
increase the state value, we have $2T_d>T_{\omega}$. We can then estimate
the maximum value of $T_d$. The partially unstable attractors typically
emerge when the total coupling strength $\varepsilon$ is relatively small
as compared with the threshold value. In this case, the  value of $T_d$
should be smaller than $T_\omega$ so the states of oscillators can become
close to the threshold. This way, oscillators can fire when receiving
pulses during the time duration of $T_\omega$. The relation between
$T_d$ and $T_\omega$ is thus
$2T_d>T_\omega>T_d$ or
\begin{equation} \label{eq_w_region}
\pi/T_d < \omega < 2\pi/T_d.
\end{equation}
To demonstrate this result, we choose a large number of parameter points in the
parameter plane $(B, \omega)$. For each point, we measure the relative
size $f_{par}$ of the regions in another parameter plane,
$(\tau,\varepsilon)$, in which partially unstable attractors arise. In
the simulations, the ranges of $\tau$ and $\varepsilon$ are set to be
$0.05\leqslant \tau \leqslant0.5$ and $0.05\leqslant \varepsilon \leqslant0.5$,
respectively, which are covered by a $46\times 46$ grid, so we have
$f_{par} \approx M_{pua}/(46\times 46)$, where $M_{pua}$ is the number
of parameter points under which period-2 partially unstable attractors exist.
For any given set of parameter values, we use $200$ random initial
conditions in the phase space to determine whether there exists any
period-2 partially unstable attractor. In particular, only when none of
the 200 initial conditions leads to such an attractor do we deem that
there is no partially unstable attractor for this parameter set. The
quantity $f_{par}$ is essentially the probability of generating partially
unstable attractors in the parameter plane $(\tau,\varepsilon)$ for any
given values of $B$ and $\omega$. Figure~\ref{fig_para}(b) shows the
dependence of $f_{par}$ on the parameters $(B,\omega)$, which gives
direct evidence that period-2 partially unstable attractors exist in
the region as predicted by Eq.~\ref{eq_w_region}, verifying the role
of the interplay between the time scale of the individual oscillators and
that of the external current in inducing such attractors.

\begin{figure*}
\centering
\includegraphics[width=\linewidth]{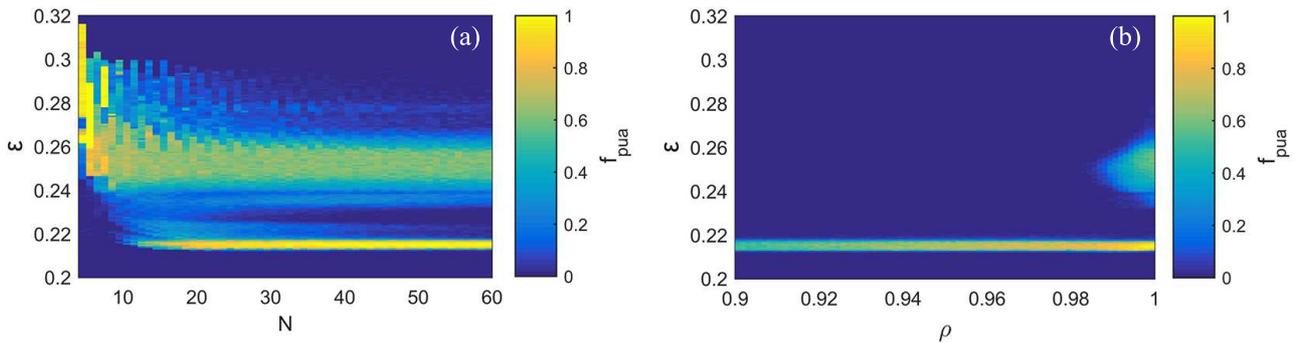}
\caption{
{\bf Existence of partially unstable attractors in globally coupled
networks of various sizes and in non-globally coupled networks}.
(a) Dependence of $f_{pua}$ on the parameters $N$ and $\varepsilon$
for $\tau=0.14$, $I=3$, $\omega=10$, and $B=1.6$ in globally coupled systems.  The non-zero value
of $f_{pua}$ demonstrates that partially unstable attractors exist in
 systems of varying sizes. (b) Dependence of $f_{pua}$
on $\rho$, the density of links, and $\varepsilon$ for $N=60$,
$\tau=0.14$, $I=3$, $\omega=10$, and $B=1.6$, which indicates that
partially unstable attractors can exist for non-globally coupled
networks of different values of the density of links. In both panels,
$f_{pua}$ denotes the fraction of initial conditions that lead to
partially unstable attractors of period 2.}
\label{fig_net_size}
\end{figure*}

\subsection{Systems of different sizes and different coupling structures}
Does the emergence of partially unstable attractors depend on the system
size $N$? To address this question we examine the $(N,\varepsilon)$ parameter
plane while fixing other parameters as $I=3$, $b=1.6$, $\omega=10,$ and
$\tau=0.14$ (the latter four parameters are the same as in Fig.~\ref{fig_pun}).
For each parameter point  in the parameter plane,
we calculate the fraction $f_{pua}$ from an ensemble of 200 random initial
conditions that lead to  period-2 partially unstable attractors.
Figure~\ref{fig_net_size}(a) shows the dependence of $f_{pua}$ on the
system size $N$ and the normalized coupling strength $\varepsilon$. We
see that the partially unstable attractors arise persistently in a wide
range of the system size.

Would the coupling structure or network topology affect the occurrence
of partially unstable attractors? For convenience, we use the density
$\rho$ of directed links, defined as $\rho=m/[N(N-1)]$, to characterize
the coupling structure of the network, where $N(N-1)$ is the number of
links in a globally coupled network and $m < N(N-1)$ directed links are
generated with each being placed between a pair of randomly selected
oscillators. We focus on the parameter plane $(\rho,\varepsilon)$ and
calculate, for each point in the plane, the fraction $f_{pua}$ of 200
random initial conditions that lead to partially unstable attractors.
Ensemble average with 30 network realizations is  used.
Figure~\ref{fig_net_size}(b) shows $f_{pua}$ versus the density of links
 $\rho$ and the normalized coupling strength $\varepsilon$. Again,
we find that partially unstable attractors are ubiquitous even when the
network topology deviates from that of global coupling.

\section{Dynamical origin of partially unstable attractors - event
based analysis.}
\subsection{Symbolic events}
There are two basic events associated with the dynamics of pulse-coupled
integrate-and-fire networks: firing and receipt (arrival) of pulses. The
events play a key role in understanding the collective dynamics such as
the source of instability of unstable attractors~\cite{zou:2010} and
as well as the classification
of multiple attractors~\cite{zou:2012}. We exploit the two events to
gain an understanding of the structure of partially unstable attractors
in terms of the formation of the stable and unstable local dynamics.

Some basic notations for the events are as follows. When a pulse from
oscillator $i$ is received by an oscillator in the network,
this event is denoted as $R_i$. A pulse will be fired (or generated)
when oscillator $j$ reaches its threshold, and this event is labeled as
$S_j$. The events occurring at different times are separated by a minus
sign. The firing events are of particular importance, where the firing
may be due to the arrival of pulses immediately, or caused by the free
dynamical evolution of the oscillator towards its threshold. In the
former case, the event is called \emph{passive firing}, where the
arrival of pulses immediately makes the state value higher than the
threshold. In the latter case, the firing occurs during the free
evolution, and is thus termed \emph{active firing}. The
effect of instantaneous perturbation on the state of an actively firing
oscillator is to cause a small change in the firing time. For example,
the sequence of events labeled as ``$R_1S_2-R_2-S_3$'' denotes three
events occurring at three different times. First, $R_1S_2$ represents
the arrival of a pulse from oscillator 1, inducing the firing of
oscillator 2. Hence $S_2$ is a passive firing event. Second, the pulse
from oscillator 2 is received ($R_2$). Third, oscillator 3 reaches its
threshold and fires: $S_3$ - an active firing event.

\subsection{Event structure of a partially unstable attractor}
\begin{figure}
\centering
\includegraphics[width=\linewidth]{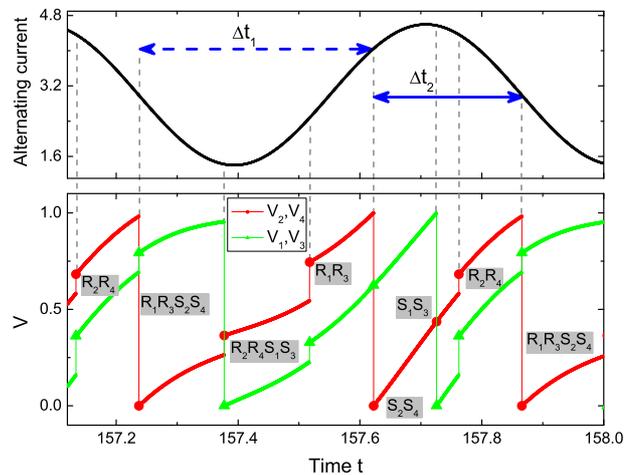}
\caption{{\bf Event analysis of a period-2 partially unstable attractor}.
For $N=4$, $\tau=0.14$, $\varepsilon=0.3$, $B=1.6$, $I=3$, and $\omega=10$,
the alternating current $I+B\cos(\omega t)$ (upper panel) and the state
variable $V$ (lower panel) of all oscillators versus time $t$ associated
with a partially unstable attractor. Oscillators 1 and 3 are synchronized,
so are oscillators 2 and 4. The states immediately after every event are
specified as red dots or green triangles. The corresponding values of the
current are indicated by the vertical dashed lines.}
\label{fig_events}
\end{figure}

How do partially unstable attractors arise and what are their event
structures? For simplicity, we study the event structure associated
with the period-2 partially unstable attractor presented in
Fig.~\ref{fig_pun}. The corresponding events are
\begin{equation} \label{event}
R_2R_4-R_1R_3S_2S_4-R_2R_4S_1S_3, \ \ \ \mbox{and} \ \ \
R_1R_3-S_2S_4-S_1S_3.
\end{equation}
Each sequence of events corresponds to the events occurring during the
time from the reset of the reference oscillator 1 to the next
reset. One property of the event structure (\ref{event}) is that
multiple oscillators become simultaneously active firing or simultaneously
passive firing at different times during one period, due to the role played
by the alternating driving current. To demonstrate this, in the upper and
lower panels of Fig.~\ref{fig_events} we show respectively the time series of
the applied current $I+B\cos(\omega t)$ and the states of the oscillators
associated with the partially unstable attractor in Fig.~\ref{fig_pun}.
The states immediately after the events are shown in dots and triangles.
The corresponding values of the alternating current are indicated by the
vertical dashed lines.

To demonstrate the role of alternative current in shaping the
structure (\ref{event}), we first consider
the time interval $\Delta t_1$ indicated in the upper panel
of Fig.~\ref{fig_events}, where the event is from passive firing
($R_1R_3S_2S_4$) to active firing ($S_2S_4$).  Oscillator 2 or 4 first
receives a pulse after a time delay $\tau$ ($R_2R_4S_1S_3$) and then two
pulses after another time delay $\tau$ ($R_1R_3$). During the time interval,
the applied current is relatively small and thus contributes little to
changing the state of oscillator 2 or 4. As a result, both oscillators 2
and 4 can generate active firing.
Now consider the time duration $\Delta t_2$, where the event is from
active firing $(S_2S_4)$ to passive firing $(R_1R_3S_2S_4)$. Oscillator 2
or 4 receives a pulse at time $\tau$ later $(R_2R_4)$, and then two pulses
at another time $t_w$ later $(R_1R_3S_2S_4)$, where $t_w < \tau$.
During $\Delta t_2$, the applied current is appreciable, so it drives
the state of oscillator 2 or 4 to a large value, generating passive firing
at the arrival of pulses from oscillators 1 and 3 $(R_1R_3S_2S_4)$.

\begin{table}[!htbp]
\parbox{7.5cm}{
\caption{ \label{Tab:N4}
A detailed description of events associated with the process
that instantaneous perturbation is applied after the 22nd reset of the
reference oscillator 1. The perturbation drives the system away from
the attractor. Here time $n$ denotes the $n$th reset of the reference
oscillator 1.}}
\begin{tabular}{l*{1}{l}r}
\hline\hline
Time $n$ \         & \  Events
   \\
\hline
18            & $R_2R_4-R_1R_3S_2S_4-R_2R_4S_1S_3$      \\
19            & $R_1R_3-S_2S_4-S_1S_3$        \\
20            & $R_2R_4-R_1R_3S_2S_4-R_2R_4S_1S_3$       \\
21            & $R_1R_3-S_2S_4-S_1S_3$        \\
\textbf{22}   & $R_2R_4-R_1R_3S_2S_4-R_2R_4S_1S_3$       \\
23     & $R_1R_3-S_4-S_2-S_3-S_1$       \\
24     & $R_4-R_2-R_3S_2S_4-R_1-R_2R_4S_1S_3$       \\
25     & $R_1R_3-S_2S_4-S_1S_3$        \\
26     & $R_2R_4-S_2S_4-R_1R_3-R_2R_4S_1S_3$       \\
27     & $R_1R_3-S_2S_4-R_2R_4S_1S_3$       \\
28     & $S_2S_4-R_1R_3-R_2R_4S_1S_3$       \\
29     & $R_1R_3-S_2S_4-R_2R_4-S_1S_3$       \\
30     & $S_2S_4-R_1R_3-R_2R_4S_1S_3$       \\
\hline
\end{tabular}
\end{table}

Intuitively, the effect of external driving on the state of each oscillator
depends on time, due to the time varying nature of the driving current.
In different time intervals, the rate of increase in the state value for
each oscillator can then be different. A strong current can induce a large
state change, which in turn can lead to passive firing, as the driving can
push the corresponding oscillator towards the threshold. However, a weak
current tends to give rise to active firings. An essential feature for
partially unstable attractors is that multiple oscillators can become
simultaneously active firing and simultaneously passive firing in
different intervals during one period of the driving. As we demonstrate
below, such an event structure is directly related to the coexistence of
stable and unstable points on the attractor.

\subsection{Understanding stable and unstable response}
We then use the event structure to understanding  the occurrence of the
stable and the unstable point for the attractor shown in Fig.~\ref{fig_pun} with event structure  (\ref{event}). 
Under the return map, the attractor is composed of two
points: $P$ and $Q$. Right after the first event sequence, the state of
the system is $P$, while $Q$ is the state of the system immediately
after the second event sequence. Here  we let the system evolve for $L=10$   periods after the system settle into the attractor. Then points $Q$ and $P$ correspond to the states of oscillators at the 21st and 22nd reset of the reference oscillator respectively.   The unstable and  stable points are
$P$ and $Q$, respectively, which can be established, as follows.

To ascertain that point $P$ is unstable, we apply perturbation to it,
which is the state of the system right after the first sequence of
events. Perturbation will affect the second sequence of events
$R_1R_3-S_2S_4-S_1S_3$. In particular, due to the perturbation,
the two simultaneously active firings $S_2S_4$ and $S_1S_3$ will be split
into four single active firings. We find that the split of the event
$S_1S_3$ is key to the emergence of the unstable local dynamics. Without
perturbation, the pulses from oscillators 1 and 3 are received
simultaneously and induce passive firing of oscillators 2 and 4 - hence
the event $R_1R_3S_2S_4$. The states of oscillators 2 and 4 right before
the $R_1R_3$ event are about 0.9821. Since the pulse strength is 0.1 and
the threshold is 1, if $\delta_3>\delta_1$, oscillators 2 and 4 will
receive one pulse first ($R_3$), which is sufficient to make them
reach the threshold. The two oscillators then receive an extra pulse
from oscillator 1 as compared with the case where there is no perturbation.
This process can cause the system to transition to a remote state in
the phase space. As a result, point $P$ is unstable. A detailed
description of the events in response to the perturbation is provided
in Table~\ref{Tab:N4}. Note that the split of $S_2S_4$ is not relevant,
because the perturbation has little effect on oscillators 2 and 4 due
to their passive firings $(R_3S_2S_4)$ during the 24th reset of the
reference oscillator, as shown in Table~\ref{Tab:N4}.

To determine that  point $Q$ is stable, we apply
instantaneous random perturbation ($\delta_1,\delta_2,\delta_3,\delta_4$)
to the states of the four oscillators, right after the reset of
oscillator 1, and examine the effect of the perturbation on the first
event sequence $R_2R_4-R_1R_3S_2S_4-R_2R_4S_1S_3$, where all four
oscillators fire passively. For example, $R_1R_3S_2S_4$ denotes the
event that oscillators 2 and 4 fire passively due to the arrival of
pulses from oscillators 1 and 3. Immediately before the occurrence of
this event, the states of oscillators 2 and 4 are slightly different
due to the perturbation. Right after the event, the states of these
two oscillators are reset to zero, effectively removing the effect of
the perturbation. Similarly, the effect of $\delta_1$ and $\delta_3$
also disappears immediately after the $R_2R_4S_1S_3$ event. Due to
the passive firings, point $Q$ is stable, i.e., the sequence of
the arrival pulses for each oscillator will not be affected by
perturbation applied to $Q$. Thus the perturbation in this case can not
change the  event structures.



\begin{table*}[t]
\parbox{10.5cm}{
\caption{\label{Tab:N60}
Examples of event structure for partially unstable attractors in
a globally coupled oscillator system of size $N=60$. Other parameters are
$\tau=0.14$, $I=3$, $\omega=10$, and $\varepsilon=0.25$. The oscillators
with a synchronized state are regarded as belonging to one group. The
number of groups determines the number of clusters in the phase space. The
events of partially unstable attractors can be represented in terms of
groups denoted as $A$, $B$, $C$, $D$, and so on. Five types of event
structures are identified for the 238 period-2 partially unstable
attractors obtained from 500 random initial conditions.}} \\
\centering
\begin{tabular}{ c | l    }
  \hline
   $i$th & Event      \\
   \hline
  1 & $R_B-R_AS_B-R_BS_A-R_A-S_B-S_A$     \\
  2 & $R_BS_C-R_AS_B-R_C-R_BS_A-R_A-S_C-S_B-R_C-S_A $     \\
  3 & $R_B-R_A-S_D-S_C-R_DS_B-S_A-R_CS_D-R_B-R_AS_C-R_DS_B-R_CS_A$     \\
  4 & $R_CS_D-R_BS_C-R_A-R_D-R_CS_B-R_BS_A-S_D-S_C-R_A-R_D-S_B-S_A$    \\
  5 & $R_CS_D-R_B-R_AS_C-S_B-R_D-R_CS_A-R_B-R_A-S_D-S_C-R_DS_B-S_A$    \\
  \hline
\end{tabular}
\end{table*}

\subsection{Event structure in larger systems.}
We then study the event structures of period-2 partially unstable attractors in large globally coupled systems. The states of  oscillators associated with
an attractor in large systems are composed of multiple clusters, analogous to the
partially unstable attractor with two clusters (Fig.~\ref{fig_pun}) for a system of size $N=4$.  Oscillators in each cluster have the same applied current and receive the same number of pulses at each arrival of pulses. The occurrence of the cluster structures is mainly due to the excitatory couplings. Suppose that some oscillators with states close to the threshold. At the arrival of some pulses at one time,  these oscillators can reset, i.e., their states become zero, inducing one cluster   for these oscillators.
One can use such cluster structures to
simplify the representation of the event structure for the whole system, where
 oscillators in the same cluster can be regarded as one group. In this way,
the event structures for different systems may have similar structures,
which can in turn be useful to understand the partially unstable
attractors in large systems.

As a concrete example,
we analyze the event structures for a system of $N=60$ globally coupled
oscillators. Due to the symmetry of the system and its high dimensionality,
a large number of period-2 partially unstable attractors can arise.
In the representation based on groups, two distinct attractors can have
the same event structure, if the corresponding groups of oscillators
exhibit the same sequence of events. However individual oscillators for
a group can be quite different, as the corresponding attractors are
different. In this way, the types of event structures can be
much fewer than the number of partially unstable attractors. For example,
from 500 random initial conditions, we obtain 238 such attractors, but
there are only 5 distinct types of event structure, as listed in
Table~\ref{Tab:N60}.

For the first example in Table~\ref{Tab:N60}, the system has two clusters
and the oscillators are organized into two groups: $A$ and $B$. Note that
for different period-2 partially unstable attractors, group $A$ or $B$
has different oscillators. For example, for one such an attractor,
group $A$ contains
27 oscillators (1, 3, 6, 7,  9, 12, 16, 17, 18, 25, 30, 31, 34, 37, 39,
40, 42, 46, 47, 48, 49, 50, 52, 53, 56, 59, and 60) and group $B$
contains all the remaining 33 oscillators. In terms of groups, we can
compare the event structures of different partially unstable attractors,
even for systems of different size. If, for the partially unstable attractor
shown in Fig.~\ref{fig_pun} for $N=4$, we assign oscillators 2 and 4 as
group $B$ and oscillators 1 and 3 as belonging to group $A$, the event
structure in (\ref{event}) is identical to that in the first example of
Table~\ref{Tab:N4}. This implies that the occurrence of partially unstable
attractors in large systems has the same dynamical mechanism as for smaller  
systems.  This is indeed the case. We consider the event $R_AS_B$ here. The state of oscillators in group $B$ just before the arrival of pulses from oscillators of group $A$ ($R_A$) is 0.9645. With perturbations on the unstable point, the oscillators of group $A$ can become 27 single active firings at slightly different times. Here each pulse of strength $0.25/59$. The first 9 arrivals of pulses can make the group $B$ passive firing, i.e.,  $0.25/99\times 9+0.9645=1.0026>1$. Thus group $B$ will receive 18, i.e., 27-9, number of extra pulses, which can make the system leave away the unstable point. This mechanism is the same as that for the partially unstable attractor shown in Fig.~\ref{fig_pun}.

\section{Conclusion and discussion}

Dissipative dynamical systems can exhibit different types of attractors.
Attractors whose neighborhoods belong completely to their basins of
attraction are the most commonly encountered type in smooth dynamical
systems.
 When the system possesses certain simple symmetry so that there
 are invariant subspaces in which there are chaotic attractors, on any such
attractor there can be a set of points that are unstable with respect to
perturbation transverse to the invariant subspace. The number of points
contained in such a set can be infinite but its measure is zero, and the
corresponding attractors have a riddled basin - a type of Milnor attractors.
Another type of Milnor attractors occurs typically in neuronal networks
that exhibit a firing or spiking behavior,
where the attractor is locally unstable but has a remote basin. Points
in the basin are attracted towards the attractor along the stable manifold,
but any random perturbation will ``kick'' the trajectory away from the
attractor. These are the unstable attractors. The main contribution of this
paper is the discovery and analysis of a novel type of attractors: partially
unstable attractors. Such an attractor is composed of two subsets: one locally
stable and another locally unstable, both of
positive measures. We have demonstrated that partially unstable attractors
can emerge in systems of excitatory pulse-coupled integrate-and-fire
oscillators subject to periodic forcing. The mechanism for the
partially unstable attractors can be understood by analyzing the dynamical
events~\cite{zou:2010,zou:2012,MT2006} leading to the generation of pulses
in the network. In particular, the event of passive firing plays a key role
in generating the locally stable set, where the effect of perturbation is
suppressed and effectively annihilated. The locally unstable set arises
due to the sensitivity of the number of arriving pulses for oscillators to
perturbation.

Our results suggest that partially unstable attractors
also persist on random networks. It is possible to study
the existence of these attractors on other types of net-
works, such as small-world networks,
or networks with communities~\cite{Newman:book}. Insofar as the essential
dynamical event structure can be identified, the possibility for partially
unstable attractors to arise can be assessed. This can be useful for
network design to achieve desired performance, e.g., for realizing
specific firing sequences for information processing. For a given network
whose structure cannot be altered, carefully controlling the periodic
forcing may lead to desired firing patterns on the network level. To
generate controlled dynamical behaviors in integrate-and-fire or more
general neuronal networks remains to be an outstanding research task at
the present.

In biological systems, information processing is often the result of
interaction between the internal dynamical state and the external
stimuli~\cite{BUONOMANO:2009}. The uncovering and understanding of
novel types of attractors in such systems can be beneficial~\cite{MPJ:2012}.
The existence of locally unstable dynamics can induce
switchings among different metastable states, which can potentially be
exploited for developing new schemes of computation~\cite{RVLHAL:2001,AB:2005,NT:2012,HYZKY:2015}. In such an application, one wishes to
generate switching dynamics that are robust to infinitesimal perturbation
but sensitive to designed forcing. The switching dynamics among
partially unstable attractors can be useful for achieving this goal. For
example, infinitesimal perturbation can be directed to locally
stable points, but forcing can be applied to locally unstable
points. At the present, to exploit partially unstable attractors to
generate robust yet sensitive switching dynamics is an open question.

\section*{Acknowledgements}
This research is supported by the National Natural Science Foundation of China (11502200, 91648101), ``The Fundamental Research Funds for the Central Universities" (No. 3102014JCQ01036), and SRF for ROCS, SEM. This research is also supported by the Aihara Project, the FIRST program from JSPS, initiated by CSTP, and CREST, JST. YCL is supported by ARO under Grant No.~W911NF-14-1-0504.

\section*{References}
\bibliographystyle{unsrt} 
\bibliography{PUA}

\end{document}